\documentclass[epj]{svjour}
\usepackage{graphicx}
\usepackage{epsfig}
\begin{document}
\title{An experimentalist's view of the uncertainties in understanding heavy element synthesis}
\author{W. Loveland
}                     
%
%
\institute{Oregon State University, Corvallis, OR 97331 USA}
\date{Received: date / Revised version: date}
%
\abstract{
The overall uncertainties in predicting heavy element synthesis cross sections are examined in terms of the uncertainties associated with the calculations of capture cross sections, fusion probabilities and survival probabilities.  Attention is focussed on hot fusion reactions.  The predicted heavy element formation cross sections are uncertain to at least one order of magnitude.
} 
\maketitle
\section{Introduction}
\label{intro}
``To deal effectively with the
doubts, you should
acknowledge their existence
and confront them straight
on, because a posture of
defiant denial is self defeating"
D. Kahneman \cite{dan}

The synthesis of new heavy nuclei is a topic that draws the attention of many scientists.  The challenge of adding to the fundamental building blocks of nature, perhaps cementing one's work  for scientific eternity, is compelling.  But the experiments and their interpretation are difficult and occasionally lead to false conclusions.  A thorough understanding of the uncertainties in measurements and theory are essential for progress.  In particular we must be able to estimate the uncertainties in both experimental data and calculated quantities from theory.  We have to able to assess the information content of a measurement and its impact on our understanding.  Measurements should test our current theories in direct and focussed ways. Theory should provide guidance as what the most important new experiments should be and how the experimental data will constrain current theories.

In this paper I review our current abilities to predict the outcome of attempts to synthesize new heavy nuclei using fusion reactions.  I try to break down the predictions into their component pieces (see below), i.e., the probability of getting the reacting nuclei to touch, the probability of having the nuclei at the contact configuration evolve inside the fission saddle point (fusion) and the probability that the fused system will survive against the destructive fission process.  I try to indicate, by comparing predictions/postdictions, with measurements related to each of these processes, the uncertainties associated with predicting the outcome of an attempt to synthesize a new heavy nucleus.

The remarkable recent progress in the synthesis of new heavy and superheavy nuclei has been made using fusion reactions.  These reactions can be divided into two prototypical classes, ``cold"  and ``hot" fusion reactions.  In ``cold" fusion reactions, one bombards Pb or Bi target nuclei with heavier projectiles (Ca-Kr) to form completely fused systems with low excitation energies ($E^\ast$=10-15 MeV), leading to a higher survival (against fission) but with a reduced probability of the fusion reaction taking place due to the larger Coulomb repulsion in the more symmetric reacting system.   In ``hot" fusion reactions one uses a more asymmetric reaction (typically involving a lighter projectile and an actinide target nucleus) to increase the fusion probability but leading to a highly excited completely fused system ($E^\ast$=30-60 MeV) with a reduced probability of surviving against fission.  

Formally, the cross section for producing a heavy evaporation residue, $\sigma$$_{\rm EVR}$, in a fusion reaction can be written as
\begin{equation}
\sigma _{EVR}(E) = \frac{{\pi {\hbar ^2}}}{{2\mu E}}\sum\limits_{\ell  = 0}^\infty  {\left( {2\ell  + 1} \right)T\left( {E,\ell } \right){P_{CN}}\left( {E,\ell } \right){W_{sur}\left( {{E^*},\ell } \right)}} 
\end{equation}
where E is the center of mass energy, and T(E,$\ell$) is the probability of the colliding nuclei to overcome the potential barrier in the entrance channel and reach the contact point.  (The term ``evaporation residue" refers to the product of a fusion reaction followed by the evaporation of a specific number of neutrons.) P$_{CN}$ is the probability that the projectile-target system will evolve from the contact point to the compound nucleus.  W$_{\rm sur}$ is the probability that the compound nucleus  will decay to produce an evaporation residue rather than fissioning.  The capture cross section is defined as
\begin{equation}
\sigma _{capt}\left( E \right) = \frac{{\pi {\hbar ^2}}}{{2\mu E}}\sum\limits_{\ell  = 0}^\infty  {\left( {2\ell  + 1} \right)T\left( {E,\ell } \right)}
\end{equation}
while the fusion cross section is
\begin{equation}
\sigma _{fus}\left( E \right) = \frac{{\pi {\hbar ^2}}}{{2\mu E}}\sum\limits_{\ell  = 0}^\infty  {\left( {2\ell  + 1} \right)T\left( {E,\ell } \right)} {P_{CN}}\left( {E,\ell } \right)
\end{equation}
The separation of the EVR cross section into three individual reaction stages (capture, fusion, survival) is motivated, in part,  by the different time scales of the processes.  For a quantitative understanding of the synthesis of new heavy nuclei, one needs to understand $\sigma _{\rm capture}$, P$_{\rm CN}$, and W$_{\rm sur}$ for the reaction system under study.	

Formally W$_{sur}$ can be written \cite{zubov,loveland,adamian,majorov,sagaidak} as 
\begin {equation}
W_{sur}\left( {{E_{c.m.}}} \right) \approx {P_{xn}}\left( {E_{CN}^*} \right)\prod_{i=1}^{x}\frac{\Gamma _{n}\left ( E_{i}^{*} \right )}{\Gamma _{n}\left ( E_{i}^{*} \right )+\Gamma _{f}\left ( E_{i}^{*} \right )}
\end{equation}
where P$_{xn}$ is the probability of emitting x (and only x) neutrons from a nucleus with excitation energy E$^{\ast}$ \cite{jackson,bobjohn}, and $\Gamma$$_{n}$ and $\Gamma$$_{f}$ are the partial widths for decay of the completely fused system by either neutron emission or fission, respectively.  (Decay of the completely fused system by charged particle emission is neglected.)  Typical procedures used to evaluate the extended product on the right hand side of equation (4) are described in \cite{zubov,loveland,majorov}.

The survival probability W$_{sur}$(E$_{c.m.}$) is of intrinsic interest.  For example, the reported cross sections for the synthesis of superheavy elements in hot fusion reactions decrease a factor of $\sim$ 3 in going from element 114 to element 118.  This behavior is attributed to the decreasing survival probability of the product nuclei as one gets further from the Z=114 shell. \cite{ogarev}.  The real situation is complicated with the fused systems starting at excitation energies, E*, of 30-50 MeV where shell effects on $\Gamma$$_{n}$/$\Gamma$$_{f}$ are not important but where dissipative effects \cite{kramers} will retard fission and ending at low excitation energies where shell effects are very important. 
	
	Several successful attempts have been made to describe the cross sections for evaporation residue formation in cold fusion reactions \cite{ssw,loveland,adamian,feng,martin}. In Figure 1a, I show some typical examples of postdictions of the formation cross sections for elements 102-113 in cold fusion reactions.The agreement between theory and experiment is impressive because the cross sections extend over six orders of magnitude, i.e., a robust agreement.  Because the values of  $\sigma_{capture}$ are well-known or generally agreed upon \cite{bock,clerc,pacheco,prokhorova,ssw}, the values of the product W$_{sur}$P$_{CN}$ are the same in most of these postdictions.  However, as seen in Figure 1b, the  values of $P_{CN}$ differ significantly in these postdictions.\cite{ssw,loveland,adamian,feng,martin,siwek2007}.  This is a clear-cut case of where a simple agreement between theory and experiment in postdicted cross sections is not sufficient to indicate a real understanding of the phenomena involved.  One must test the agreement between the components of the models involved, not just their outcome.
	
	We might ask what the overall uncertainties are in the current phenomenological models for predicting heavy element production cross sections.  Some of the leading figures in this field have commented that ``The performed analysis makes it possible to conclude that at present we are capable of calculating and predicting the values of formation cross sections of superheavy elements with Z $>$ 112 with an accuracy of two orders of magnitude at best." \cite{all}
	
	Recently one of the authors of \cite{all} produced a more nuanced response about the uncertainties in predicting new heavy nucleus formation cross sections \cite{zprivat}.  The essence of his response is that for unknown systems like symmetric reactions, the uncertainties will be several orders of magnitude.  For reactions involving $^{50}$Ti, such as the current efforts to synthesize elements 119 and 120, the uncertainties will be large.  For extrapolations to the use of targets like $^{254}$Es, the uncertainties will be at least one order of magnitude.  But for reactions involving $^{48}$Ca and ``familar" targets and products, the uncertainties will be a factor of 2. 

\section{Examination of the individual factors that influence heavy element synthesis}

\subsection{Survival Probabilities}
A recent experiment concerning survival probabilities in hot fusion reactions poses some interesting questions \cite{yanez1}.  The nucleus $^{274}$Hs was formed at an excitation energy of 63 MeV using the $^{26}$Mg + $^{248}$Cm reaction.  $^{274}$Hs has several interesting properties.  The liquid drop model fission barrier height is zero and there is a sub-shell at N=162, Z=108. In the formation reaction, P$_{CN}$ is measured \cite{itkis} to be 1.0.  The essential problem is that because of the high excitation energy, the shell effects  stabilizing the barrier are predicted \cite{naza} to be ``washed out" with a resulting fission barrier height $<$ 1 MeV. By measuring the angular distribution of the fission associated neutrons, Yanez et al. \cite{yanez1} were able to deduce a value of $\Gamma$$_{n}$/$\Gamma$$_{total}$ for the first chance fission of $^{274}$Hs (E*=63 MeV) of 0.89 $\pm$ 0.13!!  A highly excited fragile nucleus with a vanishingly small fission barrier decayed $\sim$90$\%$ of the time by emitting a neutron rather than fissioning.  Conventional calculations with various values of the fission barrier height were unable to reproduce these results.  The answer to this dilemna is the consider the effects of nuclear viscosity to retard fission\cite{kramers}.  The bottom line is that to properly evaluate survival probabilities in hot fusion reactions, one must consider both macroscopic and microscopic effects, something that is not always done.

There are other factors in the survival probabilities that are important for all types of reactions.  For example, Lu and Boilley \cite{lb} considered the effect of various fission barrier related parameters upon the calculated evaporation residue yields in the $^{208}$Pb($^{18}$O,xn) reaction.  They found collective enhancement effects, Kramers effects, and the overall fission barrier height to have the biggest effect on the calculated cross sections.  Most modern models do equally well/poorly in describing fission barrier heights for Th-Cf nuclei.  Ajanasjev et al. \cite {anatoli} found the average deviation between the calculated and known inner barrier heights was $\sim$ 0.7 MeV amongst various models.  Along the same lines, Adamian, Antonenko and Scheid \cite {aas} found differences of several orders of magnitude in evaporation residue yields depending on the mass model they used.

In summary, fission barrier heights are known to within 0.5 - 1.0 MeV, which can cause an order of magnitude error. For example in the study of the $^{238}$U($^{26}$Mg,3-4n) reaction, such an uncertainty in fission barrier heights will cause an uncertainty of a factor of 6-7 in the EVR cross sections.

\subsection{Fusion Probabilities}
No satisfactory quantitative model for the fusion probability, P$_{CN}$ exists.  Most of the effort in this area is devoted to identifying the key factors that determine the value of P$_{CN}$ for a given reacting system.  The first factor in determining P$_{CN}$ is the excitation energy of the putative compound nucleus.  Zagrebaev and Greiner \cite{one} have suggested the following ad-hoc functional form for the excitation energy dependence of P$_{CN}$
\begin{equation}
P_{CN}(E^{\ast },J)=\frac{P_{CN}^{0}}{1+\exp \left[ \frac{E_{B}^{\ast
}-E_{int}^{\ast }(J)}{\Delta }\right] }
\end{equation}
where P$_{CN}^{0}$ is the fissility dependent ``asymptotic" (above barrier)  value of P$_{CN}$ at high excitation energies, E$_{B}^{\ast}$ is the excitation energy at the Bass barrier, E$_{int}^{\ast }$(J) is the internal excitation energy (E$_{c.m.}$+Q - E$_{rot}$(J)) and $\Delta$ (an adjustable parameter)  is taken to be 4 MeV. 

Zagrebaev and Greiner \cite{one} have further suggested a Fermi function to describe the observed behavior of  P$_{CN}^{0}$at E* $\approx$ 40 MeV of
\begin{equation}
P_{CN}^{0}=\frac{1}{1+exp\left [ \frac{Z_{1}Z_{2}-\zeta }{\tau } \right ]}
\end{equation}
where $\zeta$ $\approx$ 1760 and $\tau$ $ \approx$  45.  This formula strictly applies only to cold fusion reactions but is often used for other reactions.

 Figure 2 shows a comparison of the data of Knyazheva \cite{kny} with the predictions of equations (5) and (6).  The rotational energy of the fusing system was taken from Sierk \cite{arnie}.  The agreement between the predicted and measured values of P$_{CN}$ is acceptable.  (There are alternate treatments of these data that give different values of the dependence of P$_{CN}$ upon excitation energy \cite{itkis}.)

Besides the energy dependence of P$_{CN}$, P$_{CN}$ is thought to depend on the identity of the reacting nuclei.  When two heavy nuclei collide, reaching the contact configuration (capture), they do not automatically fuse into a compound nucleus but must overcome an additional effective potential energy barrier.  The time evolution of the system of colliding nuclei will be determined by the relative values of the attractive nuclear force and the repulsive Coulomb force.  The fissility, x$_{CN}$, is a measure of the ratio of the repulsive Coulomb force and the attractive nuclear surface force for the compound nucleus. Formally
\begin{equation}
x_{CN}=\frac{\left ( \frac{Z^{2}}{A} \right )}{\left (\frac{Z^{2}}{A}\right )_{critical}}
\end{equation}
where Z and A are the atomic and mass numbers of the completely fused nucleus.  (Z$^{2}$/A)$_{crit}$ is proportional to the ratio of the nuclear surface tension coefficient to the nuclear charge density and is given as 
\begin{equation}
\left ( \frac{Z^{2}}{A} \right )_{critical}=50.883\left ( 1.-1.7826I^{2} \right )
\end{equation}
where I=(A-2Z)/A, the relative neutron excess.

It was found, upon further investigation, that the fusion process was governed not only by the parameters of the fused system (x$_{CN}$) but also by the initial binary configuration described by an ``effective" fissility, x$_{eff}$, given as 
\begin{equation}
x_{eff}=\left ( Z^{2}/A \right )_{eff}/\left ( Z^{2}/A \right )_{crit}
\end{equation}
where
\begin{equation}
x_{eff}=\frac{4Z_{1}Z_{2}/\left [ A_{1}^{1/3}A_{2}^{1/3}\left ( A_{1}^{1/3}+A_{2}^{1/3} \right ) \right ]}{\left ( Z^{2}/A \right )_{critical}}
\end{equation}
x$_{eff}$ best describes the necked shapes near the mass asymmetric contact configuration while x$_{CN}$ applies to shapes without a neck characteristic of the completely fused system.

Blocki et al. \cite{blocki} found that best scaling value to describe various aspects of the fusion process,  x$_{m}$, could be constructed from a combination of x$_{CN}$ and x$_{eff}$.  They found that using $ x_{m}^{Blocki}$= (2/3)x$_{CN}$ + (1/3)x$_{eff}$  was the best choice of a scaling variable. 

du Rietz et al. \cite{dur} made an extensive survey of P$_{CN}$ in a large number of fusing systems.  They thought that perhaps some fissility related parameter would be the best way to organize their data on P$_{CN}$ and its dependence of the properties of the entrance channel in the reactions they studied.  They found the best fissility-related scaling variable that organized their data was $x_{m}^{du Rietz}$= 0.75x$_{eff}$ + 0.25x$_{CN}$, a different combination than Blocki had found.  They also called their scaling variable, the mean fissility.  These differing and perhaps contradictory scaling variables are symptomatic of our understanding  of P$_{CN}$ and the phenomena being represented.

A summary of the current measured values of P$_{CN}$ for E*=40-50 MeV is shown in figure 3.  The mean ``du Rietz" fissility, $x_{m}^{du Rietz}$, is a generally adequate gross scaling value for P$_{CN}$.  However the tenfold dispersion in P$_{CN}$ values around $x_{m}^{du Rietz}$ = 0.65 indicates there are other factors, besides $x_{m}^{du Rietz}$ that influence P$_{CN}$.  Another caveat about figure 3 is that the measured values of P$_{CN}$ are generally $>$ 0.10, a condition dictated by the experimental techniques used to measure P$_{CN}$.  The P$_{CN}$ values that are more relevant for the synthesis of the heaviest nuclei may be $<$ 0.01 and are difficult to measure.

\subsection{TDHF Calculations}

In principle, Time Dependent Hartree-Fock (TDHF) calculations of the fusion of two colliding nuclei ought to be able to predict capture cross sections and fusion/quasifission cross sections.  These calculations are difficult, time consuming and may need comparison with experiment to properly interpret them.  Recently Wahkle et al. \cite{wahkle} made a pioneering study of the $^{48}$Ca + $^{238}$U reaction.  The capture cross sections predicted by their TDHF calculations agreed with measured capture cross sections \cite{shen} within $\pm$ 20 $\%$.  In addition they were able to predict the ratio of fusion to capture cross sections, $\frac{\sigma _{fus}}{\sigma _{capt}}$, as 0.09 $\pm$ 0.07 at 205.9 MeV and 0.16 $\pm$ 0.06 at 225.4 MeV in agreement with \cite{shen} who measured ratios of 0.06 $\pm$ 0.03 and 0.14 $\pm$ 0.05, respectively.  Additional information about the role of the orientation of the colliding nuclei at contact and the role of nuclear shell structure in the collision was obtained.  Whether TDHF calculations can become a predictive tool for heavy element synthesis remains to be seen.

\subsection{Capture Cross Sections}

The capture cross section is, in the language of coupled channel calculations, the ``barrier crossing" cross section.  It is the sum of the quasifission, fast fission, fusion-fission and fusion-evaporation residue cross sections.  The latter cross section is so small for the systems studied in this work that it is neglected.  The barriers involved are the interaction barriers and not the fusion barriers.  There are several models for capture cross sections \cite{ssw,bian,zaggyweb,wang}.  Each of them has been tested against a number of measurements of capture cross sections for reactions that, mostly, do not lead to the formation of the heaviest nuclei.  In general, these models are able to describe the magnitudes of the capture cross sections with 50$\%$ and the values of the interaction barriers within 20$\%$.  Amongst the many methods for estimating capture cross sections, the use of coupled channels calculations \cite{zaggyweb} appears to do the best job of predicting the capture cross sections.

 In Table 1, I compare the results of a set of calculations  (using \cite{zaggyweb}) of the capture cross sections for reactions that synthesize heavy nuclei with the measured cross sections.  I chose to compare the predictions and data at excitation energies that are relevant.  Good agreement between the measured and calculated values of the cross sections occurs for all reactions. 
 
\section{Summary}

Each of the factors that affect the production of new heavy nuclei using complete fusions reactions has significant uncertainties.  Some of these individual factors are uncertain to an order of magnitude or more.  The inclusion of macroscopic factors such as nuclear viscosity in calculating hot fusion survival probabilities for the synthesis of the heaviest elements seems mandatory and is included in \cite{zaggyweb} but not in all predictions. (This factor will increase the survival probability. )  Phenomenological prescriptions for hot fusion cross sections that describe the experimental data without including this factor probably contain compensating errors in their treatment of the cross sections.  The summary of current predictions of the heights of heavy element fission barriers indicates an mean uncertainty of about 0.7 MeV in these estimates.  Since hot fusion cross sections involve multiple chance fission (n=3-5), the propagation of errors can lead to further uncertainties.  Underlying all this is a disagreement upon the values of the nuclear masses for the heaviest nuclei.  The nuclear masses give us the shell corrections which are the fission barriers for the heaviest nuclei.  As Pei et al. \cite{naza} have shown, the damping of the shell corrections with increasing excitation energy differs for each nucleus, a fact that is rarely taken into account in estimates of production cross sections.  Unfortunately fission barrier heights are not pure observables in that the deduced values from experiments depend upon assumptions in reaction modeling.  

The problem of understanding the fusion probability, P$_{CN}$,  is a difficult one.  Experimentalists are starting to give gross phenomenological descriptions but even these are incomplete.  TDHF calculations are being made but nothing resembling a serious theory exists.  Some success has occurred in Langevin calculations of quasi-fission processes and calculations of P$_{CN}$ in the $^{48}$Ca + $^{248}$Cm reaction by Zagrebaev and Greiner \cite{JPG}

All of these comments are written from the perspective of an experimentalist not an expert in nuclear theory.  As experimentalists we seek to use the simple tools available to us, such as \cite{zaggyweb},  to make the needed estimates to allow us to efficiently carry out experiments or to select which experiments are worth performing.  We would also hope that we could gain some insight into the physics of the processes involved

To our colleagues in nuclear theory, we experimentalists plead for help.  Imagine you/we are trying to  synthesize a new chemical element or nuclide.  Somebody's phenomenology/theory says the cross section is 100 fb.  With modern apparatus, that might be a production rate of a few atoms per year.  (Or is it 1 atom per 100 years?)

\begin{acknowledgement}
This work was supported, in part, by the Director, Office of Energy Research, Division of Nuclear Physics of the Office of High Energy and Nuclear Physics of the U.S. Department of Energy under Grant DE-FG06-97ER4102
\end{acknowledgement}

\begin{figure}
\epsfxsize 8.4cm
\epsfbox{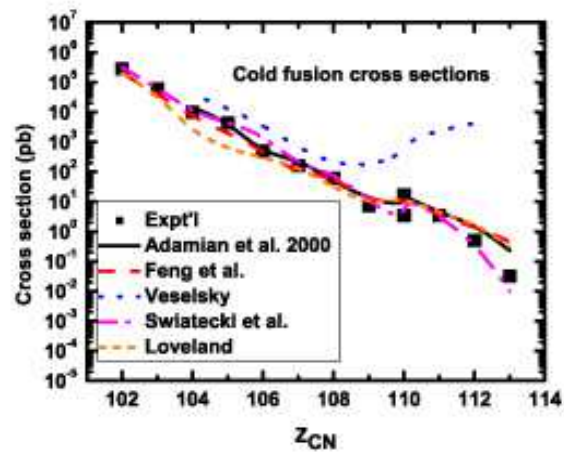}
\epsfbox{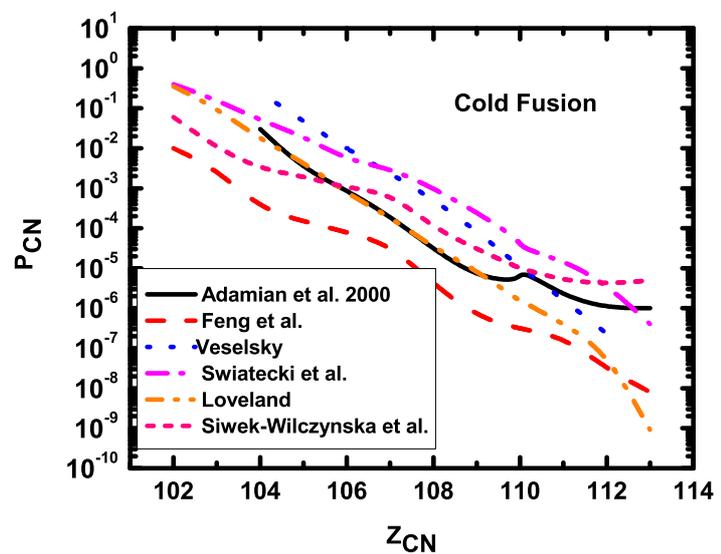}
\caption{\label{fig1} (Color online)(a). Typical predictions of the formation cross sections of elements 102-113 using cold fusion reactions.  (b)  Comparison of predictions of P$_{CN}$ for these cold fusion reactions}
\end{figure}

\begin{figure}
\begin{minipage}{20pc}
\begin{center}
\includegraphics[width=17pc]{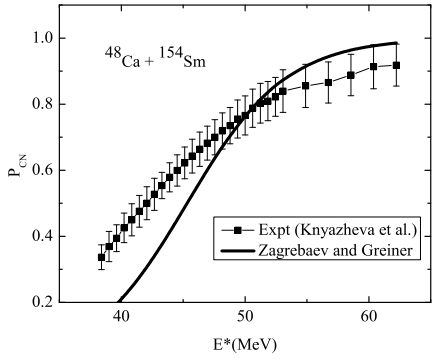}
\end{center}
\caption{\label{fig2}Comparison of measured \cite{kny} and predicted \cite{one} values usign equations [5] and [6] of P$_{CN}$ for the reaction of $^{48}$Ca with $^{154}$Sm.}
\end{minipage} 
\end{figure}

\begin{figure}
\begin{minipage}{28pc}
\begin{center}
\includegraphics[width=28pc]{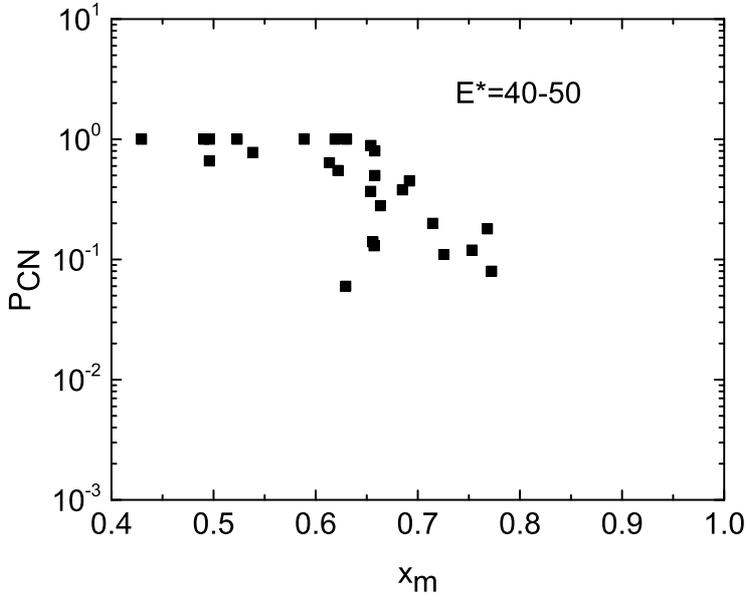}
\end{center}
\caption{\label{fig3}Measured values of P$_{CN}$ for E*=40-50 MeV}
\end{minipage} 
\end{figure}

\begin{table}[b]
\begin{center}
\vspace{0.5cm}
\caption{Measured and predicted capture-fission cross sections}
\begin{tabular}{ccccccccc}
Proj.&Target&CN&E$_{\rm c.m.}$ &$E^\ast$&Expt.&$\sigma_{calc}$\cite{zaggyweb}&$\sigma_{calc}$/$\sigma_{meas}$\\
        &            &      &             (MeV)   &    (MeV) & (mb)&           (mb)                                      &                                                              \\
\hline
$^{36}$S&$^{208}$Pb&$^{244}$Cf&153.9&40.&363.\cite{Y}&201.4&0.55 \\
$^{30}$Si&$^{238}$U&$^{268}$Sg&133.9&40.&21\cite{N}&18.1&0.86\\
$^{58}$Fe&$^{208}$Pb&$^{264}$Hs&245.5&40.&200.\cite{I}&165.4&0.83\\
$^{26}$Mg&$^{248}$Cm&$^{274}$Hs&122.3&40.&9\cite{itkis}&17.9&2.0\\
$^{34}$S&$^{238}$U&$^{274}$Hs&151.6&40.&20.\cite{N}&13.3&0.67\\
$^{48}$Ca&$^{238}$U&$^{286}$Cn&199.2&40.&100.\cite{I}&125.2&1.25\\
$^{48}$Ca&$^{244}$Pu&$^{292}$Fl&196.3&35&25.\cite{I}&38.6&1.55\\
$^{48}$Ca&$^{248}$Cm&$^{296}$Lv&202.3&35&25\cite{I}&56.1&2.24\\
\label{table1}
\end{tabular}
\end{center}
\end{table}

\end{document}